\documentclass[12pt]{article}%
\usepackage{graphics}%
\def\preprint%
#1{\thispagestyle{empty}~\newline\vspace*{-22.65mm}%
\begin{flushright}%
\begin{tabular}{l} #1%
\end{tabular}%
\end{flushright}%
\vspace{1cm}}%
\begin{document}%
\markright{\hfil Particle detectors, geodesic motion, and the
equivalence principle}%
\title{\bf \LARGE Particle detectors, geodesic motion, and the
equivalence principle}%
\author{Sebastiano Sonego$^*$ and Hans Westman$^\dagger$\\[2mm]%
{\small \it \thanks{\tt sebastiano.sonego@uniud.it}~Universit\`a
di Udine, Via delle Scienze 208, 33100 Udine, Italy}\\%
{\small \it \thanks{\tt hawe@fy.chalmers.se}~Department of
Astronomy and Astrophysics,}\\%
{\small\it Chalmers University of Technology, 41296 G\"oteborg,
Sweden}}%

\date{{\small October 29, 2003; \LaTeX-ed \today }}%
\maketitle%
\begin{abstract}%
It is shown that quantum particle detectors are not reliable
probes of spacetime structure.  In particular, they fail to
distinguish between inertial and  non-inertial motion in a general
spacetime.  To prove this, we consider detectors undergoing
circular motion in an arbitrary static spherically symmetric
spacetime, and give a necessary and sufficient condition for the
response function to vanish when the field is in the static vacuum
state.  By examining two particular cases, we show that there is
no relation, in general, between the vanishing of the response
function and the fact that the detector motion is, or is not,
geodesic. In static asymptotically flat spacetimes, however, all
rotating detectors are excited in the static vacuum.  Thus, in
this particular case the static vacuum appears to be associated
with a non-rotating frame. The implications of these results for
the equivalence principle are considered.  In particular, we
discuss how to properly formulate the principle for particle
detectors, and show that it is satisfied.%
\\%

\end{abstract}%
\def\g{{\mbox{\sl g}}}%
\def\Box{\nabla^2}%
\def\d{{\mathrm d}}%
\def\L{{\cal L}}%
\def\SIZE{1.00}%
\def\lab{\label}%
\newpage%

\section{Introduction}%
\lab{sec:intro}%

A particle detector in Minkowski spacetime remains unexcited when
the field is in the vacuum state, if it moves along a geodesic,
i.e., along a straight line in space at a constant speed
\cite{bd}. This property is often regarded as a quantum
counterpart of the law of inertia --- quantum detectors and
classical mechanics experiments identify the same class of
preferred reference frames\footnote{Hereafter, by {\em observer\/}
we mean a future-directed, differentiable timelike curve in
spacetime. A {\em reference frame\/} is a congruence of
observers.} \cite{smolin}.  By a straightforward application of
the equivalence principle, one would then expect that in an
arbitrary spacetime, a detector in geodesic motion should not
become excited, provided the field is in the vacuum state.
However, this statement, as well as its converse --- that a
non-inertial detector in an arbitrary spacetime is somewhat
analogous to an accelerated detector in Minkowski spacetime, and
should therefore get excited\footnote{This is often used as a
heuristic explanation for particle creation by a gravitational
field.} ---, is trivially false, unless formulated more precisely.
Indeed, in a non-stationary spacetime there is no meaningful
notion of a ``vacuum state'', because particle creation takes
place \cite{bd}. In other cases one can have inequivalent vacua
\cite{fulling,letaw-pfautsch}, and the opposite problem arises.
This happens already in Minkowski spacetime, where a uniformly
accelerated particle detector in the so-called Rindler vacuum does
not become excited, although it does so in the usual Minkowski
vacuum \cite{bd,grishchuk}.  Then, the
notion of a vacuum becomes ambiguous.%

Nevertheless, one may define a vacuum state directly in terms of
detectors response, considering their behaviour in all possible
frames, for each possible state of the field.  If there happens to
be a state, for which there exists at least one reference frame in
which the detectors carried by all observers are not excited, then
one can identify it with a vacuum.  Conversely, knowing in advance
the vacuum state, one can find the frames in which detectors do
not get excited.  Thus, a null response can be used in two
different ways, in order to identify both a vacuum state and one
or more preferred frames associated with it.\footnote{This
operational definition of a vacuum is not directly equivalent, in
general, to the one based on canonical field quantisation
\cite{letaw-pfautsch}.  See, however, reference \cite{ottewill}
for an attempt to reconcile the two definitions, based on taking
into account the spurious effects due to the detector structure.
The operationalistic approach could be criticised by noticing that
``particles'' are not fundamental objects in quantum field theory,
which is in fact about fields.}%

We can now reformulate the statement above in a more precise way.
Suppose that in a given spacetime one can find a state of the
field in which detectors carried by one particular geodesic frame
are not excited.  Then, one expects that the same should happen
for {\em all\/} geodesic frames, and {\em only\/} for them.  One
aim of the present note is to disprove this claim by exhibiting
two counterexamples.  In doing so, we shall also show that quantum
detectors are not reliable probes of spacetime structure. This
fact, in turn, will help elucidating the status
of the equivalence principle in quantum field theory.%

Our strategy is based on the fact that, in a static spacetime, one
can define a state in which static quantum detectors are not
excited --- a static vacuum.  Of course, in general this has
nothing to do with the state in which an inertial detector does
not become excited.  However, in the particular case of a
so-called ultrastatic spacetime \cite{ultra}, a static frame is
also geodesic.  It is then sufficient to find ultrastatic
spacetimes such that, when the field is in a static vacuum, either
there are non-geodesic detectors which also do not get excited, or
there are geodesic detectors which do get excited. Taken together,
these two types of behaviour show that there is no correlation
whatsoever between the response of a detector and the fact that
its motion is, or is not, geodesic.%

The article is organised as follows.  We restrict our attention to
the case of static spherically symmetric spacetimes.  In the next
section we define a static vacuum and present some background
calculations about the response function of a quantum detector
that moves on a circular orbit.\footnote{The response of rotating
detectors was also considered in other papers for different
purposes than ours.  Such previous analysis have been performed in
order to gain a deeper understanding of the detector response
\cite{ottewill,letaw},  to investigate the properties of
alternative vacuum states \cite{davies,brasilians}, and to explore
the possibility for experimental tests of Unruh-like effects
\cite{bell}.} In section \ref{sec:bounds} we establish a necessary
and sufficient condition for the response function to vanish in
the static vacuum.  In section \ref{sec:qrf} we show that in a
static, asymptotically flat spacetime, rotating detectors do not
become excited when the field is in a static vacuum iff their
worldlines coincide with those of the static observers. Hence, in
this case detectors identify the same state of non-rotation of
classical mechanical experiments. In general, however, they fail
to do so.  This happens, e.g., in Einstein's static universe,
where there are non-geodesic motions for which the no-detection
condition is satisfied, as shown in section \ref{subsec:einstein}.
Conversely, in section \ref{subsec:bulge} we describe a class of
spacetimes for which the condition is violated by detectors that
move freely along circular orbits --- hence, with geodesic
worldlines. Section \ref{sec:concl} contains a discussion about
the validity of the equivalence principle in quantum field theory,
and some concluding remarks.%

\section{Preliminaries}%
\lab{sec:response}%
\setcounter{equation}{0}%

In the following, we consider a real scalar field $\phi$, obeying
the field equation\footnote{We work in units with $c=\hbar=1$, and
choose the metric signature to be $+2$. Tensor indices from the
beginning of the Latin alphabet, $a$, $b$, $c$, ..., run from 0 to
3.  The conventions for the curvature tensors are the same adopted
in reference \cite{wald}.}%
\begin{equation}%
{\g\,}^{ab}{\nabla\!}_a{\nabla\!}_b\,\phi-\xi\,R\,\phi-\mu^2\,\phi=0%
\label{E:fieldeq}%
\end{equation}%
in a spacetime $({\cal M},\g_{ab})$, where $\xi$ and $\mu$ are
constants and $R$ is the Ricci curvature scalar.%

\subsection{Vacuum state}%
\label{subsec:vacuum}%

Let $f_I$ be positive-norm functions that, together with their
complex conjugates $f_I^\ast$, form a complete set in the space of
solutions of (\ref{E:fieldeq}), and let $|0\rangle$ be the state
such that $\hat{a}_I|0\rangle=0$, where the annihilation
operators $\hat{a}_I$ are defined through the decomposition%
\begin{equation}%
\hat{\phi}(x)=\sum_I\left(f_I(x)\,\hat{a}_I+
f_I(x)^\ast\,\hat{a}_I^\dagger\right)%
\label{annihilation}%
\end{equation}%
of the field operator $\hat{\phi}$.  Then, the response function
of a DeWitt monopole detector for $\phi$ in the state $|0\rangle$
is \cite{bd}%
\begin{equation}%
{\cal R}(E;\gamma)=\lim_{T\to
+\infty}\frac{\Theta(E)}{2\,T}\int_{-T}^{T}
\d\tau\int_{-T}^{T}\d\tau'\,{\rm e}^{-{\rm
i}\,E(\tau-\tau')} G^+\left(x(\tau),x(\tau')\right)\;.%
\label{E:response}%
\end{equation}%
Here $\gamma$ denotes the detector worldline, expressed by the
coordinate functions $x^a(\tau)$ of the detector proper time
$\tau$, $\Theta$ is the step function, and%
\begin{equation}%
G^+\left(x,x'\right)=\langle
0|\,\hat{\phi}(x)\,\hat{\phi}(x')\,|0\rangle=\sum_I
f_I(x)\,f_I(x')^\ast%
\label{E:D+}%
\end{equation}%
is the Wightman function.  Physically, ${\cal R}(E;\gamma)$ is the
probability per unit time that the detector be excited from its
ground state to the energy eigenstate with energy $E$. This
transition is usually regarded as corresponding to the detection
of a quantum of the field with energy $E$, although such an
interpretation is not free from ambiguities \cite{interpretation}.
If the modes $f_I$ have positive frequency with respect to
$\gamma$,
i.e., if%
\begin{equation}%
\int_{-\infty}^{+\infty}{\rm d}\tau\,{\rm e}^{-{\rm i}\,E\,\tau}
f_I(x(\tau))%
\label{pos-freq}%
\end{equation}%
is non-zero only for negative values of $E$, then ${\cal
R}(E;\gamma)=0$ and $|0\rangle$ is recognised as a vacuum
state according to the operational definition given in
section \ref{sec:intro}.%

\subsection{Modes and frequency spectrum}%
\label{subsec:modes}%

We work in a static, spherically symmetric spacetime with the
metric
\begin{equation}%
\g=-A(\rho)^2\d t^2+\d\rho^2+r(\rho)^2\left(\d\theta^2
+\sin^2\theta\,\d\varphi^2\,\right)\;,%
\label{sph-metric}%
\end{equation}%
where $A$ and $r$ are two positive functions, that we leave
unspecified for the moment.  A convenient set of positive-norm
solutions of the field equation (\ref{E:fieldeq}) is then given
by modes of the form%
\begin{equation}%
f_{lm}^{(\sigma)}(\rho,\theta,\varphi\,|\,\omega)\,{\rm e}^{-{\rm
i}\,\omega t}\;,%
\label{E:modes}%
\end{equation}%
with $\omega>0$.  In general, $\omega$ belongs to a spectrum which
has both a discrete and a continuous part, and the label $\sigma$
accounts for a possible degeneracy of the parameters $\omega$,
$l$, and $m$ \cite{frolov-novikov}.  These are positive-frequency
modes with respect to static observers.  Hence, the corresponding
vacuum state is such that static detectors register no particles.
We shall refer to this as the {\em static vacuum\/} (also known
as the ``Boulware vacuum'' \cite{boulware}).%

Because of spherical symmetry, we can separate variables and
write straightforwardly%
\begin{equation}%
f_{lm}^{(\sigma)}(\rho,\theta,\varphi\,|\,\omega)=
\frac{(-1)^m}{r}\,\chi_{l}^{(\sigma)}(\rho\,|\,\omega)
\left(\frac{2l+1}{4\pi}\,\frac{(l-m)!}{(l+m)!}\right)^{1/2}
P_l^m(\cos\theta)\,{\rm e}^{{\rm i}\,m\,\varphi}\;,%
\end{equation}%
where $P_l^m$ are the associated Legendre functions and $l$, $m$
are integer numbers with $l\geq 0$ and $-l\leq m\leq l$ (see,
e.g., reference \cite{arfken}).  The functions
$\chi_l^{(\sigma)}(\rho\,|\,\omega)$ are then solutions of the
differential equation%
\begin{equation}%
\frac{\d^2\chi_l^{(\sigma)}}{\d x^2}+
\left(\omega^2-\frac{1}{r}\frac{\d^2 r}{\d x^2}-\xi\,A^2\,R
-A^2\,\mu^2-A^2\,\frac{l(l+1)}{r^2}\right)\chi_l^{(\sigma)}=0\;,%
\label{E:eigenvalue}%
\end{equation}%
where the variable $x$, defined by%
\begin{equation}%
\d x=\frac{1}{A(\rho)}\,\d\rho\;,%
\label{dxdr}%
\end{equation}%
is a generalised Regge-Wheeler ``tortoise'' coordinate \cite{sm}.
Formally, (\ref{E:eigenvalue}) is a time-independent
one-dimensional Schr\"odinger equation for a non-relativistic
particle of mass $1/2$ and energy $\omega^2>0$ in an external
field with potential energy%
\begin{equation}%
V_l(x)=\frac{A(\rho)}{r(\rho)}\frac{\rm d}{{\rm
d}\rho}\left(A(\rho)\,\frac{{\rm d}r(\rho)}{{\rm
d}\rho}\right)+\xi\,A(\rho)^2\,R(\rho)
+A(\rho)^2\,\mu^2+A(\rho)^2\,\frac{l(l+1)}{r(\rho)^2}\;,%
\label{V}%
\end{equation}%
where we have transformed the first term on the right hand side
using (\ref{dxdr}), and $\rho$ must be regarded as a function of $x$,
obtained by integrating (\ref{dxdr}) and inverting the result.
Thus, equation (\ref{E:eigenvalue}) represents an eigenvalue
problem: For an arbitrary value $l\geq 0$, solving
(\ref{E:eigenvalue}) means not only to find a function
$\chi_l^{(\sigma)}(\rho\,|\,\omega)$, but also to determine the
spectrum ${\cal S}_l$ of the frequencies $\omega\in {\cal S}_l$
compatible with the assigned $l$. The total
frequency spectrum is then%
\begin{equation}%
{\cal S}=\bigcup_{l=0}^{+\infty}{\cal S}_l\;.%
\label{S}%
\end{equation}%

In general, ${\cal S}_l$ (hence $\cal S$) is the union of a
discrete and a continuous part.  Qualitatively, the structure of
the spectrum ${\cal S}_l$ can be determined by analysing the
potential energy $V_l$ given by (\ref{V}), just as one does
traditionally in quantum mechanics.  Indeed, the possible values
$\omega\in {\cal S}_l$ coincide with the square root of the energy
eigenvalues for the time-independent Schr\"odinger equation
(\ref{E:eigenvalue}).\footnote{Of course, only the non-negative
part of the quantum mechanical spectrum leads to real values of
$\omega$.}  As it is well-known, values of the total energy
corresponding to an unbounded classical one-dimensional motion
with potential energy $V_l$, belong to the continuous part of the
quantum mechanical spectrum.  On the other hand, values of the
total energy that belong to the discrete part of the spectrum,
must necessarily correspond to classically bound motions
\cite{gottfried}.%

\subsection{Response function on circular orbits}%
\label{subsec:response}%

Let us consider detector worldlines of the particular type%
\begin{equation}%
\left.\begin{array}{l}%
t(\tau)=\alpha\,\tau\\ %
\rho(\tau)=\mbox{const}\\ %
\theta(\tau)=\pi/2\\ %
\varphi(\tau)=\Omega\,t(\tau)=\alpha\,\Omega\,\tau%
\end{array}\right\}\,,%
\label{E:worldline}%
\end{equation}%
which correspond to uniform circular motions. Here $\Omega$ is a
constant, and%
\begin{equation}%
\alpha=\left(A^2-\Omega^2\,r^2\right)^{-1/2}%
\end{equation}%
by normalisation of the four-velocity.  Thus, the possible values
of $\Omega$ for an orbit at a given $\rho$ are constrained by the
inequality\footnote{This condition can also be re-expressed as
$|\Omega|\widetilde{r}<1$, where $\widetilde{r}=r/A$ is the
circumferential radius in the corresponding optical geometry
\cite{sm}.  The parameter $\widetilde{r}$ plays the role of radius
of gyration for relativistic dynamics \cite{gyration}.}%
\begin{equation}%
|\Omega|<A(\rho)/r(\rho)\;.%
\label{E:limit}%
\end{equation}%
Substituting the parametric equations (\ref{E:worldline}) for the
worldline into the modes (\ref{E:modes}), then using equations
(\ref{E:response}) and (\ref{E:D+}) one gets, after performing the
integrals in $\tau$ and $\tau'$ and taking the limit, the response
function for a detector which moves along a circular orbit
identified by the parameters $\rho$ and $\Omega$:%
\begin{eqnarray}%
{\cal
R}(E;\rho,\Omega)&=&\displaystyle{\frac{\Theta(E)}{\pi\,r^2}\sum_\sigma
\sum_{l=0}^{+\infty}\sum_{m=-l}^l 2^{m-1}(2\,l+1)\,
\frac{(l-m)!}{(l+m)!}\,\frac{\Gamma\left(\displaystyle{\frac{l+m+1}{2}}\right)}
{\Gamma\left(\displaystyle{\frac{l-m}{2}+1}\right)}\,
\cos^2\frac{\left(l+m\right)\pi}{2}}\nonumber\\%
&&\int_0^{+\infty} \d\omega\,{\lambda\,}_l(\omega)\,
\delta\left(E+\alpha(\rho)\,\omega- m\,\alpha(\rho)\,\Omega\right)
\left|\chi_l^{(\sigma)}(\rho\,|\,\omega)\right|^2\;,%
\label{E:specresp1}%
\end{eqnarray}%
where we have used the explicit expression for $P_l^m(0)$
\cite{abramowitz}. In this expression, ${\lambda\,}_l(\omega)$ is
a non-negative generalised function which accounts for the
frequency spectrum, such that ${\cal
S}_l=\left\{\omega\left|\;{\lambda\,}_l(\omega)\neq
0\right.\right\}$. In general, ${\lambda\,}_l$ is a sum of delta
functions (discrete part of the spectrum) and step functions
(continuous part).%

\section{No-detection condition}%
\lab{sec:bounds}%
\setcounter{equation}{0}%

The detector with worldline (\ref{E:worldline}) remains unexcited
iff ${\cal R}(E;\rho,\Omega)=0$ for all values of $E$. Because of
the occurrence of the step function $\Theta(E)$ in equation
(\ref{E:specresp1}), this condition is equivalent to the vanishing
of the integral on the right-hand side of (\ref{E:specresp1}) for
all positive $E$.  Since all terms of the sum in
(\ref{E:specresp1}) are non-negative, the detector is not excited
iff each of them, taken separately, is equal to zero. Neglecting
coefficients different from zero, we find that the
no-detection condition amounts to asking that%
\begin{equation}%
\cos^2\frac{\left(l+m\right)\pi}{2}\int_0^{+\infty}{\rm
d}\omega\,{\lambda\,}_l(\omega)\,\delta\left(E+\alpha(\rho)\,\omega-
m\,\alpha(\rho)\,\Omega\right)\,
\left|\chi_l^{(\sigma)}(\rho\,|\,\omega)\right|^2=0\;,%
\label{E:nosum}%
\end{equation}%
$\forall E>0$, and for all possible combinations of $l$, $m$, and
$\sigma$. The cosine coefficients vanish when $l+m$ is odd, hence
(\ref{E:nosum}) leads to a non-trivial no-detection condition only
for even values of $l+m$. Since the measure function
${\lambda\,}_l$ is, by definition, non-zero when $\omega\in {\cal S}_l$,
this condition is%
\begin{equation}%
\delta\left(E+\alpha(\rho)\,\omega- m\,\alpha(\rho)\,\Omega\right)\,
\left|\chi_l^{(\sigma)}(\rho\,|\,\omega)\right|^2=0\;,%
\label{E:noint}%
\end{equation}%
$\forall E>0$, and for all possible combinations of $l$,
$\omega\in {\cal S}_l$, $m$, and $\sigma$, with $l+m$ even. One
may worry about the possibility of having degenerate values of
$\rho$, where $\chi^{(\sigma)}_l(\rho\,|\,\omega)=0$ for all
possible combinations of $l$, $\omega\in {\cal S}_l$, and
$\sigma$. However, the set of all possible values of $\rho$ for
which this can happen must have zero measure, because one requires
completeness of the set of modes.  Thus, for almost all $\rho$,
the set%
\begin{eqnarray}%
{\cal P}_\rho=\{(\omega,l)\,|\,\omega\in {\cal S}_l,\,l\geq
0,\,\exists\,\sigma\ \mbox{such that}\
\chi^{(\sigma)}_{l}(\rho\,|\,\omega)\neq 0\}%
\end{eqnarray}%
is nonempty. We conclude that the detector at $\rho$ does not click
iff the delta function in (\ref{E:noint}) is zero for all
$(\omega,l)\in {\cal P}_\rho$, i.e., iff%
\begin{equation}%
E+\alpha(\rho)\left(\omega-m\,\Omega\right)\neq 0\;,%
\end{equation}%
$\forall E>0$, for all pairs $(\omega,l)\in {\cal P}_\rho$, and
values of $m$ such that $l+m$ is even. (See also reference
\cite{davies}.) Since $E$ must be positive and $\alpha$ is
positive by definition, this condition
is equivalent to%
\begin{equation}%
m\,\Omega\leq\omega\;,%
\label{E:mOmega<omega}%
\end{equation}%
again for all $(\omega,l)\in {\cal P}_\rho$, and $m$ such that $l+m$
is even. Since $\omega\geq 0$, the case $m=0$ always trivially
satisfies (\ref{E:mOmega<omega}), so one can express the
no-detection condition more conveniently as
$-\Omega^{-}_\rho\leq\Omega\leq\Omega^{+}_\rho$,
where%
\begin{equation}%
\Omega^{-}_\rho:=\inf\left\{\omega/(-m)\,|(\omega,l)\in {\cal
P}_\rho,\ l+m\ \mbox{even},\ m<0\right\}%
\end{equation}%
and%
\begin{equation}%
\Omega^{+}_\rho:=\inf\left\{\omega/m\,|(\omega,l)\in {\cal
P}_\rho,\ l+m\ \mbox{even},\ m>0\right\}\;.%
\end{equation}%
It is easy to check that%
\begin{equation}%
\Omega^{-}_\rho=\Omega^{+}_\rho=\Delta_\rho:=
\inf\left\{\omega/l\,|(\omega,l)\in {\cal P}_\rho\right\}\;,%
\label{Delta}%
\end{equation}%
so the necessary and sufficient no-detection condition for a
specific value of $\rho$ becomes
\begin{equation}%
|\Omega|\leq\Delta_\rho\;.%
\label{nodetcond}%
\end{equation}%

If $\Delta_\rho=0$, the no-detection condition reduces to
$\Omega=0$:  Only detectors belonging to the non-rotating frame do
not get excited.  In general, however, there could be an entire
band of values of $\Omega$ around $\Omega=0$ for which no
detection takes place.  The value of $\Delta_\rho$, which gives
the exact width of this band, can be computed by (\ref{Delta}),
once the structure of the frequency spectrum is known. In
particular, for each given value of $\rho$ one must know the
possible values of $\omega/l$.  This, in turn, requires the
knowledge of ${\cal S}_l$, which is determined by finding the
eigenvalues of the Schr\"odinger operator in (\ref{E:eigenvalue})
for the given value of $l$.  For a concrete example of this
calculation, see section \ref{subsec:einstein} below.%

\section{Quantum non-rotating frame}%
\lab{sec:qrf}%
\setcounter{equation}{0}%

In the particular case of an asymptotically flat spacetime, we
have%
\begin{equation}%
\lim_{\rho\to +\infty}r(\rho)=+\infty\;,%
\end{equation}%
\begin{equation}%
\lim_{\rho\to +\infty} A(\rho)=\lim_{\rho\to +\infty} r'(\rho)=1\;,%
\end{equation}%
and%
\begin{equation}%
\lim_{\rho\to +\infty} A'(\rho)=\lim_{\rho\to +\infty} r''(\rho)=
\lim_{\rho\to +\infty} R=0\;,%
\end{equation}%
where $A'$, $r'$, and $r''$ are derivatives of the functions $A$ and
$r$.  It follows that%
\begin{equation}%
\lim_{\rho\to +\infty}V_l(\rho)=\mu^2\;,%
\end{equation}%
for any value of $l$. This implies that the continuous part ${\cal
S}_{\rm cont}$ of the spectrum is ${\cal S}_{\rm cont}=
\,]\mu,+\infty[\,$, independent of the value of $l$.%

Since $\cal S$ has a continuous part, there is no constraint on
$l$ in correspondence to the values of $\omega$.  (This can be
understood by the quantum mechanical analogy, where the continuous
part of the spectrum corresponds to scattering states, for which
every value of $l$ is allowed.)  Then, $\inf \{\omega/l\}=0$,
because for any given value of $\omega$ in ${\cal S}_{\rm cont}$,
there is no upper bound to the possible values of $l$, so one can
trivially consider the limit $l\to +\infty$.  Hence, $\Delta_\rho=0$,
which selects just the static frame.%

\section{Detector response and geodesic motion}%
\lab{sec:einstein}%
\setcounter{equation}{0}%

In its roughest version, the claim mentioned in the Introduction
--- that a particle detector should not get excited in a vacuum
state, if and only if its motion is geodesic --- is trivially
false. Indeed, a static detector does not get excited when the
field is in the static vacuum defined in section
\ref{subsec:vacuum}, and yet its motion is not geodesic for
non-constant $A$.  One might see this, however, as a drawback of
that particular vacuum state rather than evidence for the
incorrectness of the claim.  After all, the same happens in
Minkowski spacetime, where a uniformly accelerated detector may
not get excited, provided the field is in the so-called Rindler
vacuum \cite{bd,grishchuk}.%

However, if the spacetime is not just static, but {\em
ultrastatic\/} --- which in our case simply amounts to having
$A\equiv 1$ \cite{ultra} ---, then static observers are also
geodesic.  Hence, the static vacuum is certainly appropriate, and
one might expect that any other geodesic detector, and
only a geodesic one, should not get excited in it.%

We now exhibit two counter-examples to this claim.  Taken
together, they show that, in general, there exists no relation
between zero/non-zero response function and geodesic/non-geodesic
motion.  Since, in the next section, we want to relate this result
to a quantum version of the equivalence principle, it is important
to ensure that no violation of it can take place already at the
classical level. In fact, it turns out \cite{SonegoFaraoni} that,
among  all possible field equations of the form (\ref{E:fieldeq}),
only the one with $\xi=1/6$ (the so-called conformal coupling
between field and curvature) satisfies Einstein's equivalence
principle.\footnote{We follow reference \cite{will} for the
terminology about the various possible formulations of the
principle.}  Thus, in this section we shall restrict ourselves  to
considering a conformally coupled scalar field.%

\subsection{Einstein's static universe}%
\lab{subsec:einstein}%

The inequality (\ref{nodetcond}) has interesting consequences in
spacetimes with finite spatial sections \cite{davies}. Consider
the conformally coupled scalar field in Einstein's static
universe, $A=1$ and $r(\rho)=a\,\sin(\rho/a)$, where $a$ is a
positive constant.
In this case, the frequency spectrum is discrete,%
\begin{equation}%
\omega_n=\frac{1}{a}\,\left(\left(n+1\right)^2
+\mu^2\,a^2\right)^{1/2}\;,%
\end{equation}%
with $n\geq l$, and the functions $\chi_l(\rho\,|\,\omega_n)$ are
proportional to the Gegenbauer polynomials $C_{n-l}^{l+1}(r/a)$
\cite{ford}.  Then we have ${\cal S}_{l}=\{\omega_l,\omega_{l+1},...\}$,
and%
\begin{eqnarray}%
\lefteqn{\Delta_\rho=\frac{1}{a}\,\inf\left\{\left.\left(\left(n+1\right)^2+
\mu^2\,a^2\right)^{1/2}/l\;\right|\;
n=0,1,\ldots ;\; l=0,1,\ldots,n ;\; C_{n-l}^{l+1}(r/a)\neq
0\right\}}\nonumber\\%
&=&\frac{1}{a}\,\inf\left\{\left.\left(\left(n+1\right)^2+
\mu^2\,a^2\right)^{1/2}/n\;\right|\;n=0,1,\ldots ;\;
C_0^{n+1}(r/a)\neq 0\right\}=\frac{1}{a}\;,%
\end{eqnarray}%
because $C_0^{n+1}(r/a)\equiv 1$.  Hence, the no-detection condition
turns out to be%
\begin{equation}%
|\Omega|\,a\leq 1%
\label{E:cond}%
\end{equation}%
for any value of $\rho$.  The same conclusion can be obtained by
noticing that the no-detection condition is equivalent to%
\begin{equation}%
n\,|\Omega|\leq\omega_n%
\end{equation}%
for all values of $n$, so%
\begin{equation}%
|\Omega|\,a\leq \left(\left(1+\frac{1}{n}\right)^2
+\frac{\mu^2\,a^2}{n^2}\right)^{1/2}%
\end{equation}%
for all $n$.  The tighter constraint corresponds to $n\to
+\infty$, and gives again (\ref{E:cond}).  Notice that our
condition (\ref{E:cond}) is at variance with inequality (16)
in reference \cite{davies}.%

Since the detector must travel at a speed smaller than the speed
of light, we also have from (\ref{E:limit}) that $|\Omega|<1/r$.
On a great circle $r=a$, this inequality coincides with the
no-detection condition (\ref{E:cond}).  Therefore, a detector with
constant speed on a great circle (and hence in geodesic motion)
does not become excited (see also reference \cite{davies}). This,
at first sight, seems to support the conjecture, because motion at
a constant speed along a great circle is geodesic.  But consider
instead a detector on a circular orbit that is not a great circle,
$r<a$. Such an orbit does not correspond to geodesic
motion.\footnote{In any ultrastatic spacetime, a spacetime
geodesic corresponds to motion along a spatial geodesic at a
constant speed.}  The no-detection condition (\ref{E:cond}),
however, is still satisfied, provided $|\Omega|$ is smaller than
$1/a$.  Thus, for sufficiently small speeds, the detector is not
excited, although its motion is non-geodesic.  We can also
conclude from this result that the  detector fails to reveal the
state of non-rotation.%

Interestingly, all this follows directly from a general theorem
due to Chmielowski, who has shown that the vacuum states
associated with two global complete commuting timelike Killing
vector fields coincide \cite{chmielowski}.  Indeed, in the
Einstein static universe, $\partial_t$ and $\partial_t +
\Omega\,\partial_\varphi$ are two such fields, when $\Omega$ is a
constant such that $|\Omega|<1/a$.  This implies that detectors
with four-velocity parallel to $\partial_t +
\Omega\,\partial_\varphi$, where $|\Omega|<1/a$, do not get
excited when the field is in the static vacuum, in agreement with
the analysis presented above.%

\subsection{Asymptotically flat ultrastatic spacetime}%
\lab{subsec:bulge}%

When the function $r(\rho)$ possesses local maxima or minima,
there are circular geodesics $\rho=\mbox{const}$ on the spatial
sections $t=\mbox{const}$.  These can be easily visualised in
terms of the occurrence of throats and bulges in the embedding
diagram of equatorial planes \cite{sm}. Motion at constant
speed along these circles corresponds to some particular
timelike geodesics in spacetime.  From the analysis in
section \ref{sec:qrf}, it follows that in an asymptotically flat
spacetime, detectors on these geodesics do get excited in the
static vacuum.\footnote{The ``conspiratorial'' possibility that
the modes $\chi_l^{(\sigma)}$ could all vanish just for these
values of $\rho$ can be excluded by considering a spacetime in
which the bulge, or throat, is actually a finite portion of a
flat cylinder.  In this case the modes should all vanish within
a finite interval of values of $\rho$, which is excluded by
completeness.} From this particular class of examples,
one realises that not only detectors do not reveal the spacetime
structure; they do not even reveal the geometry of
three-dimensional space (either ordinary or optical).%

\section{Discussion}%
\lab{sec:concl}%
\setcounter{equation}{0}%

We have presented both an example in which a non-geodesic detector
does not become excited (section \ref{subsec:einstein}), as well
as one where a geodesic detector has a non-vanishing response
function (section \ref{subsec:bulge}).  Thus, in general, there is
no correlation between the vanishing of the response function of a
particle detector in a vacuum state, and the fact that the
detector worldline is, or is not, a geodesic in spacetime.%

With a little hindsight, it is easy to realise that there really
was little reason to believe in the existence of such correlation.
The response function does not depend directly on local geometry
but on {\em global\/} field modes. That there is a nice
coincidence in Minkowski spacetime between geodesic motion and
non-detection of quanta when the field is in the vacuum state,
should be regarded as a very special result.  In a generic static
spacetime, one can define a vacuum in which any static detector is
not excited.  If the spacetime is ultrastatic, any static detector
is also geodesic, so there is a state in which detectors belonging
to one particular geodesic frame do not get excited. In the
degenerate case of Minkowski spacetime, there are infinitely many
such static frames, related to each other by Lorentz
transformations, all defining the same vacuum, by Chmielowski's
theorem \cite{chmielowski}.  Together, these frames contain all
the geodesic motions in Minkowski spacetime, and this explains why
the detector remains in the ground state iff it moves along a
geodesic.  But this circumstance is accidental, and due to the
fact that the high degree of symmetry of Minkowski spacetime
allows one to identify the classes of static and geodesic frames.
To generalise this, by claiming that the no-detection condition
should always identify geodesic frames, is incorrect, as shown
by the counter-examples of section \ref{sec:einstein}.%

The same sort of degeneracy that we have in Minkowski spacetime
also occurs in the Einstein static universe, as we saw in section
\ref{subsec:einstein}.  There, all frames rotating with a
sufficiently small angular velocity correspond to a null response.
This might be understood, heuristically, thinking of the particle
detection, and of the corresponding field excitation, as a
resonance phenomenon,\footnote{We are grateful to Rickard Jonsson
for pointing this out to us.} an interpretation suggested by the
form of the inequality (\ref{E:mOmega<omega}).  The quantity
$\omega\,r/m$ is the phase speed, along the detector trajectory,
of the field modes characterised by $\omega$ and
$m$.\footnote{Notice that this is not the same quantity as the
{\em component\/} of the phase velocity along the trajectory.  If
$v^i_{\rm ph}$ are the coordinate components of the phase velocity
($i$ and $j$ denote spatial indices, running from 1 to 3), and
$k_i$ is the wave-number one-form, then the phase speed in the
direction defined by the unit vector  $e^i$ is $\left(k_i\,v_{\rm
ph}^i\right)/\left(k_j\,e^j\right)$.  The component of the phase
velocity along the same direction is instead
$h_{ij}\,e^i\,v^j_{\rm ph}$, where $h_{ij}$ is the metric  of
three-space.}  For a discrete spectrum, there is a lower bound  in
$\omega/m$ (equal  to $1/a$ in the Einstein universe), so a
detector whose angular velocity is below that limit fails to
excite any mode and remains in its ground state. However, in an
asymptotically flat spacetime the spectrum has a continuous part
and the quantity $\omega/m$ can be arbitrary low, so a detector
moving with any arbitrarily low speed can always excite a field
mode.  This removes the degeneracy mentioned above.  In fact, the
quantum vacuum in this case happens to be associated with a
non-rotating frame.%

The whole analysis of the present paper shows that the
local behaviour of detectors in a curved background does differ
from the one in Minkowski spacetime.  It seems therefore, at first,
that using particle detectors one can invalidate the equivalence
principle.  Such a conclusion is, however, incorrect.  Einstein's
equivalence principle applies only to experiments performed within
a spacetime region $\cal O$ that satisfies two requirements:%
\begin{description}%
\item[{\rm (i)}]  $\cal O$ is very small with respect  to the
typical scales associated with curvature;%
\item[{\rm (ii)}]  The physical conditions of a non-gravitational
nature in $\cal O$ are the same that one would have within a
similar region in Minkowski spacetime.%
\end{description}%
Since we consider point-like detectors, the detection process
takes place locally, and requirement (i) is satisfied.  However,
the detector necessarily interacts with the field, which ``feels''
the entire spacetime through its quantum state --- definitely a
global concept.  Hence, requirement (ii) does not apply, and local
field observables in $\cal O$ behave, in general, differently than
they would in a similar region of Minkowski spacetime.  Although
the detector behaviour is at variance with what one would expect
on the basis of the equivalence principle, one cannot speak of a
true violation, because the requirements under which the principle
itself can be applied are not fulfilled.\footnote{See reference
\cite{grishchuk} for similar considerations about uniformly
accelerated detectors.} In short, it simply makes no sense to ask
whether the particle detectors considered so far satisfy the
equivalence principle.%

There is a similarity between this issue and an old problem
concerning the behaviour of electric charges in a gravitational
field.  The Lorentz-Dirac equation in a curved spacetime contains
a term that accounts for possible effects of radiation
backscattering off the curvature \cite{dw-b}.  This also seems to
violate the equivalence principle, but what is actually violated
are, again, the very conditions for testing it, because the
behaviour of a charge cannot be regarded as local, since it
depends on its entire electromagnetic field.  In particular, the
radiation produced by the charge can be backscattered off the
curvature, and affect the charge itself at later times.  Thus,
experiments involving radiation reaction, even if performed within
a small neighbourhood of the charge --- hence satisfying (i) ---,
are in general sensitive to the effects on it of curvature in
regions that are far-away in spacetime.  Therefore, such
experiments do not meet requirement (ii).  One can remove this
objectionable feature by suitably adapting the experimental
setting, though.  For example, one can shield the charge enclosing
it within a small metal box, thus preventing it from being
affected by any radiation coming from distant regions --- in
particular, radiation generated through the backscattering of the
charge's own field off spacetime curvature.  The behaviour of the
charge will then be the same as it would have been within a
similar box in Minkowski spacetime. This should not appear as an
implementation of the equivalence principle by brute force.
Indeed, the very reason why the principle itself is interesting at
all, is in order to check whether position invariance and Lorentz
invariance still hold true, locally, in a curved spacetime
\cite{will}.  Requirement (ii) is then necessary in order to
exclude trivial violations of these properties.%

Similarly, one can conceive experiments involving a particle
detector in which requirement (ii) is also fulfilled.  Enclose the
detector within a perfectly reflecting box (i.e., impose the
boundary condition $\phi\equiv 0$ on the walls of the box). The
modes inside the box are now completely independent of the
``external world,'' so the detector is, in fact, isolated from the
field outside.  In addition to this shielding, let us also assume
that the coefficient $\xi$ in the field equation (\ref{E:fieldeq})
is equal to $1/6$, so that no violation of the equivalence
principle takes place already at the classical level
\cite{SonegoFaraoni}.  Then, the modes inside the box will be
indistinguishable from those that one would have within a similar
box in Minkowski spacetime, provided the box is sufficiently
small. The response function will then be the same in the two
cases \cite{peres}.  Hence, it appears that Einstein's equivalence
principle holds also for particle detectors, once requirement
(ii) is duly fulfilled.%

\section*{Acknowledgements}

We wish to thank an anonymous board member for suggesting several
improvements.  SS is grateful to Antony Valentini for asking a
question that stimulated this investigation, and to Tony Rothman
for bringing reference \cite{davies} to his attention. He also
gratefully acknowledges warm hospitality from the Department of
Astronomy and Astrophysics at Chalmers University, where part of
this work
was done.%

%
\end{document}